\documentclass[a4paper,fleqn]{cas-dc}

\usepackage[numbers]{natbib}
\usepackage{lmodern}  
\usepackage{amsmath}  
\usepackage{xcolor}   
\usepackage{listings}
\usepackage{subcaption}
\usepackage[super]{nth}
\def\tsc#1{\csdef{#1}{\textsc{\lowercase{#1}}\xspace}}
\tsc{WGM}
\tsc{QE}

\usepackage{xcolor,colortbl}

\usepackage{todonotes}

\usepackage{listings}
\usepackage{xcolor}

\lstdefinelanguage{C}{
  morekeywords={int, float, double, char, void, if, else, while, for, return, 
    break, continue, switch, case, default, static, struct, typedef, enum, const},
  keywordstyle=\color{blue}\bfseries,
  commentstyle=\color{gray}\ttfamily,
  stringstyle=\color{orange},
  morecomment=[l]//,
  morecomment=[s]{/*}{*/},
  morestring=[b]",
  sensitive=true
}

\definecolor{codegray}{rgb}{0.95,0.95,0.95} 
\definecolor{codeblue}{rgb}{0,0,0.6}
\definecolor{codegreen}{rgb}{0,0.5,0}
\definecolor{codered}{rgb}{0.6,0,0}

\usepackage{longtable}
\usepackage{booktabs}
\usepackage{underscore}

\usepackage{array}
\usepackage{multirow}
\newcolumntype{P}[1]{>{\raggedright\arraybackslash}m{#1}}
\newcolumntype{M}[1]{>{\centering\arraybackslash}m{#1}}  

\begin{document}
\let\WriteBookmarks\relax
\def\floatpagepagefraction{1}
\def\textpagefraction{.001}

\shorttitle{CI for Dynamic Resource Management in HPC}    

\shortauthors{P. Sandås et al.}  

\title [mode = title]{A Test Taxonomy and Continuous Integration Ecosystem for Dynamic Resource Management in HPC}

\author[1]{Petter Sandås}[orcid=0009-0009-2028-7719]
\author[1]{Íñigo Aréjula-Aísa}
\author[1]{Sergio Iserte}[orcid=0000-0003-3654-7924]
\ead{sergio.iserte@bsc.es}
\cormark[1]
\cortext[1]{Corresponding author}
\author[1]{Antonio J. Peña}[orcid=0000-0002-3575-4617]

\affiliation[1]{organization={Barcelona Supercomputing Center (BSC)},
            city={Barcelona},
            country={Spain}}





\begin{abstract}
High-performance computing (HPC) systems are increasingly exploring dynamic resource management and malleable MPI applications to better adapt to heterogeneous architectures, fluctuating workloads, and energy constraints. However, the correctness of the libraries that support these techniques is often evaluated through ad hoc experiments that can be difficult to reproduce and maintain. This article introduces methodology for testing dynamic resource management frameworks that combines a taxonomy of tests for MPI malleable libraries with an HPC-oriented continuous integration (CI) ecosystem. The taxonomy structures functional and non-functional tests at both component-integration and system levels. The CI ecosystem instantiates this taxonomy in a containerized virtual cluster enabling automated validation. The approach is instantiated and evaluated using the Dynamic Management of Resources (DMR) framework as a representative case study. Results show that the proposed methodology improves early fault detection, simplifies maintenance under evolving dependencies, and transfers to other malleability solutions that expose analogous primitives for initialization, readiness checking, and reconfiguration.
\end{abstract}



\maketitle

\section{Introduction}\label{sec:intro}
Dynamic resource management (DynRM) is becoming increasingly relevant as HPC systems face more heterogeneous architectures, fluctuating workloads, and tighter energy and throughput constraints. Malleability based on Message Passing Interface (MPI) allow applications to  adjust their resource footprint at runtime, offer a powerful mechanism to exploit these dynamics, but they also introduce intricate interactions between applications, runtimes, and resource managers that are difficult to validate and maintain.

Despite significant progress in malleability mechanisms and DynRM frameworks, validating their correctness and performance still often relies on ad hoc practices: developers rebuild bespoke stacks, reconfigure batch systems, and run manual experiments to check that resizing logic, data redistribution, and error handling behave as expected. As these solutions evolve and are integrated with multiple runtimes and schedulers, the lack of structured testing and automation becomes a bottleneck for sustainable development.

In this work, the generic DynRM testing and continuous integration (CI) approach is instantiated and evaluated using the Dynamic Management of Resources (DMR) framework, a representative DynRM solution in this space. Over the years, DMR has evolved into a mature research software asset and has been integrated into a diverse set of production-oriented scientific applications, ranging from computational mechanics and environmental modelling to molecular dynamics and genomic sequencing~\cite{Vazquez15d,iserte2025dmr_sweg,Caviedes2023,eupilot2021eu,rojek_parallelization_2015,iserte_study_2020,martinez2013dynamic,iserte_dynamic_2018}. This adoption, together with ongoing work on process-management mechanisms (e.g., Proteo, DPP, ULFM)~\cite{martin-alvarez_dynamic_2023,iserte_towards_2025,huber2024designprinciplesdynamicresource,huber_bridging_2025,iserte2025dmr_dpp_oar,garcia_hints_2014,de_rosso_empowering_2025} and resource-management back ends (e.g., Slurm extensions, OAR, and hybrid CPU–QPU workflows)~\cite{iserte2025dmr_dpp_oar,10.1007/978-3-031-90203-1_34,iserte_resource_2025,iserte_mpi_2025,rocco2025dynamic}, illustrates both the versatility of DMR and the complexity of maintaining correctness across its many integration points.

This work aims to address these challenges by providing a systematic view of how DynRM and malleable MPI solutions can be tested and continuously validated in realistic HPC environments. Section~\ref{sec:motivation} motivates the need for such an approach by reviewing the DynRM landscape and highlighting recurring malleability patterns across existing frameworks. Building on this, Section~\ref{sec:methodology} introduces a generic structure for tests and an HPC-oriented CI ecosystem, while later sections instantiate and evaluate the approach in DMR.

The remainder of this paper is organized as follows. Section~\ref{sec:motivation} presents the motivation and problem statement. Section~\ref{sec:related} reviews related work around DynRM and CI methodologies in HPC. Section~\ref{sec:methodology} introduces the proposed test structure and CI ecosystem for dynamic resource management. Section~\ref{sec:implementation} presents a use case where this approach is instantiated using a malleability framework. Section~\ref{sec:evaluation} evaluates this use case under different scenarios. Section~\ref{sec:discussion} discusses how the methodology extends to other DynRM solutions, and Section~\ref{sec:conclusions} concludes the paper.

\section{Motivation and Problem Statement}\label{sec:motivation}
DynRM denotes mechanisms that allow a running job to change the amount and type of resources it uses while it is executing, instead of maintaining a fixed allocation decided at submission time. In HPC systems, DynRM enables jobs to adapt to heterogeneous nodes or changing performance conditions without stopping and restarting the application, for instance, by expanding jobs when spare nodes are available, or shrinking them under load or energy constraints.

\begin{figure}
    \centering
    \includegraphics[width=0.6\linewidth]{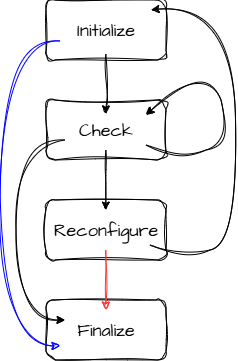}
    \caption{Basic malleability workflow: the application initializes, periodically checks for reconfiguration, and, when needed, adjusts resources and processes before continuing execution.}
    \label{fig:malleability-flow}
\end{figure}

Figure~\ref{fig:malleability-flow} illustrates the basic workflow of a dynamic execution. After the initial setup, the application may periodically reaches synchronization points where a potential reconfiguration is evaluated (\textit{check}).  If a change is requested (for example, adding resources and creating additional processes or shrinking the allocation), a \textit{reconfiguration} step is performed.
The reconfiguration involves a re-initialization step (for example, respawning processes or creating additional ones), after which the execution resumes and the cycle of checks and possible reconfigurations is repeated.

Although the execution can terminate from any stage, the blue arrow from initialization to finalization represents a non-dynamic scenario in which no reconfiguration can occur. In contrast, the red arrow from reconfiguration to finalization highlights an undesirable case where the system incurs the overhead of a resize only to terminate immediately afterwards.

Within this broader DynRM space, MPI malleability focuses on applications that use MPI and can adjust their process layout at runtime. A malleable MPI application can add or remove MPI processes during execution, in coordination with the resource management system and the MPI runtime, while preserving correct communication and data distribution. This requires tightly coupled interactions between the application, a malleability library, the MPI process manager, and the batch scheduler, and small changes in any of these layers can affect the correctness of resize decisions, process management, and data movement.

\begin{figure}
    \centering
    \includegraphics[width=0.8\linewidth]{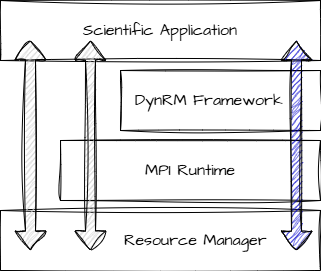}
    \caption{Technology stack for an MPI malleable application, showing interactions between the application, DynRM framework, MPI runtime, and resource manager.}
    \label{fig:malleability-stack}
\end{figure}

Figure~\ref{fig:malleability-stack} represents the technology stack of an MPI malleable application.
Particularly, the dynamic scenario is represented by the blue arrow which enables the interconnection among the scientific applciation, the dynamic resource management framework, the MPI runtime, and eventually, the resource manager system.

Notice that the scheme is compatible with applications that are dicrectly submitted to the resource manager, as well, as MPI applications submitted to the resource amangener (gray arrows).

Figure~\ref{fig:malleability-stack} represents the technology stack of an MPI malleable application. The dynamic scenario is highlighted by the blue arrow, which shows how the scientific application, the dynamic resource management framework, the MPI runtime, and the resource management system interact to support malleability.

The scheme is compatible both with applications (MPI-enabled or not) submitted directly to the resource manager, as indicated by the gray arrows.

Existing DynRM frameworks and malleability solutions have demonstrated that process-level malleability is feasible and beneficial in production-oriented codes.
However, their validation still largely relies on bespoke experimental setups, manual test campaigns, and ad hoc scripts that rebuild stacks and reconfigure batch systems for each study. This practice turns results hard to reproduce, increases the cost of maintaining DynRM libraries under changing MPI and resource manager system (e.g., Slurm) releases, and offers reduced support for early detection of subtle regressions.

Our work addresses these gaps by introducing a test taxonomy and CI ecosystem tailored to DynRM libraries and MPI malleability frameworks. It leverages the similar semantic structure of many MPI malleability solutions to define a structured space of both functional and non-functional tests at component-integration and system level, aligned with a standard malleability workflow (initialization, check, and reconfiguration). Furthermore, it presents a CI ecosystem that instantiates this structure in a containerized, virtual cluster that executes tests against MPI runtimes and resource managers.

The ideal test taxonomy and CI ecosystem should be \emph{framework-agnostic}. In this context, the focus is on existing malleability frameworks that naturally support both API-level and integration testing. Since works based on SCR extensions~\cite{Lemarinier2016}, EasyGrid AMS~\cite{Ribeiro2013}, and ReSHAPE~\cite{Sudarsan2007} do not publicly provide an API or examples on how to implement malleability, the discussion concentrates instead on PCM malleability~\cite{ElMaghraoui2007}, AMPI~\cite{Gupta}, Flex-MPI~\cite{Martin2013}, Elastic MPI~\cite{Compres2016}, ULFM-based malleability~\cite{ulfm}, DPP~\cite{huber2024designprinciplesdynamicresource}, and Proteo~\cite{martin-alvarez_proteo_2024}. From the perspective developed in this work, these frameworks share a common pattern: they support \textbf{system-level tests} that exercise end--to--end malleability scenarios with multiple initialization, check, and reconfiguration cycles, enabling both \textbf{functional} and \textbf{non-functional} assessment of their behavior.

\paragraph{Process Checkpointing and Migration (PCM)}
In PCM-based applications~\cite{ElMaghraoui2009}, the malleability workflow maps naturally to the workflow stages: \emph{initialization} is handled by \texttt{PCM\_MPI\_Init} and \texttt{PCM\_Init}; \emph{check} is performed through periodic \texttt{PCM\_Status} calls that assess whether a migration or resize should occur, while respecting guards and no-op behavior; and \emph{reconfiguration} is driven by \texttt{PCM\_Store} and \texttt{PCM\_Reconfigure}, which coordinate resource reallocation, communicator reshaping, and data redistribution. A DynRM-oriented CI workflow targeting PCM must therefore exercise these primitives explicitly and integrate with PCM’s native scheduling mechanisms so that dynamic resource allocation and process migration are tested under realistic conditions.

\paragraph{Adaptive MPI}
In AMPI, the malleability workflow appears through a combination of initialization, handoff points, and runtime-driven migration. \emph{Initialization} covers AMPI configuration and memory management (e.g., \texttt{isomalloc}), which implicitly registers data for migration. The \emph{check} phase corresponds to points where the application hands control to the runtime through calls such as \texttt{AMPI\_Migrate} with \texttt{MPI\_Info} hints (e.g., \texttt{ampi\_load\_balance=sync}), acting as reconfiguration requests at synchronization points. \emph{Reconfiguration} is then carried out by CHARM++, which migrates virtual ranks, redistributes registered data, and reshapes the process layout. A DynRM-oriented CI workflow for AMPI must therefore integrate AMPI’s load-balancing and migration mechanisms (via CHARM++) with the batch system (e.g., Torque/Maui as in~\cite{Prabhakaran2015}) so that migration and resize decisions are exercised automatically.

\paragraph{Flex-MPI}
In Flex-MPI, the workflow is similarly structured around initialization, monitoring, and runtime-driven adaptation. \emph{Initialization} comprises MPI and Flex-MPI setup, including calls such as \texttt{XMPI\_Get\_wsize}, data registration (\texttt{XMPI\_Register}), and retrieval of shared data before computation. The \emph{check} phase is executed each iteration by \texttt{XMPI\_Monitor\_init} and \texttt{XMPI\_Eval\_reconfiguration}, where performance data are collected and evaluated to decide on reconfiguration. \emph{Reconfiguration} is handled by the Flex-MPI runtime, which adapts the process set and redistributes registered data; shrink events are exposed via \texttt{XMPI\_Get\_process\_status}, marking processes as \texttt{XMPI\_REMOVED}. A CI ecosystem targeting Flex-MPI should integrate its monitoring and reconfiguration mechanisms (built on MPICH) with the ad hoc resource manager so that different growth and shrink scenarios can be exercised reproducibly.

\paragraph{Elastic MPI}
In Elastic MPI, malleability is made explicit through adaptive initialization, probing, and communicator updates. \emph{Initialization} is handled by \texttt{MPI\_Init\_adapt}, which distinguishes between initially launched and joining ranks and performs the appropriate setup. The \emph{check} phase occurs at periodic calls to \texttt{MPI\_Probe\_adapt}, where processes test whether an adaptation is requested and whether they should participate. \emph{Reconfiguration} is implemented by \texttt{MPI\_Comm\_adapt\_begin} and \texttt{MPI\_Comm\_adapt\_commit}, within which data redistribution is performed, covering resource reallocation, process layout updates, and application-level data movement. A CI ecosystem for Elastic MPI should integrate these adaptive mechanisms with the Slurm/MPICH stack so that automated tests can drive join and adapt cycles through the native resource manager.

\paragraph{User-Level Failure Mitigation (ULFM)}
In ULFM-enabled applications, failure-aware primitives can be repurposed to support malleability-like shrink and reconfiguration. \emph{Initialization} involves switching the default error handler (e.g., on \texttt{MPI\_COMM\_WORLD}) to a non-fatal handler and preparing the application to interpret ULFM error codes such as \texttt{MPIX\_ERR\_PROC\_FAILED}, \texttt{MPIX\_ERR\_REVOKED}, and \texttt{MPIX\_ERR\_PROC\_FAILED\_PENDING}. The \emph{check} phase occurs when communication operations return these codes or when the application explicitly invokes failure-detection routines, thereby deciding whether reconfiguration is needed. \emph{Reconfiguration} is realized through \texttt{MPIX\_Comm\_revoke} and especially \texttt{MPIX\_Comm\_shrink}, which construct a new communicator without failed (or intentionally removed) processes; combined with standard MPI process-management mechanisms for expansion, these operations implement shrink/expand while leaving data redistribution to the application. A CI ecosystem for ULFM-style malleability should couple ULFM-capable MPI runtimes with fault-injection or process-control facilities to exercise shrink and recovery patterns.

\paragraph{MPI Sessions with Dynamic Process Sets (DPP)}
In applications using DPP, the generic malleability workflow maps cleanly to the sessions and process-set interfaces. \emph{Initialization} comprises starting a session, discovering the initial process set, and creating the working communicator (e.g., \texttt{MPI\_Session\_init}, process-set lookup, \texttt{MPI\_Group\_from\_session\_pset}, \texttt{MPI\_Comm\_create\_from\_group}), followed by application-specific setup. The \emph{check} phase is realized by periodic, non-blocking interaction with the process-set operation interface: ranks initiate changes via \texttt{MPI\_DPP\_psetop\_nb} and query progress with \texttt{MPI\_DPP\_psetop\_query} while normal computation proceeds. \emph{Reconfiguration} takes place when a non-null operation is reported: the application exchanges data, updates the current process set, disconnects the old communicator, and recreates a new one, covering resource reallocation, process-layout changes, and application-managed data redistribution. A CI ecosystem targeting DPP should integrate an implementation of DPP with the MPI sessions mechanisms in Open MPI and the associated resource manager so that automated tests can drive process-set operations (requests, publication, cancellation).

\paragraph{Proteo}
In Proteo-based applications, the malleability workflow is embodied in the Malleability Module (MaM)~\cite{martin-alvarez_mam_2025} and its source/target split and data-registration mechanisms. The \emph{initialization} phase combines standard MPI configuration with \texttt{MAM\_Init}, which selects the source or target path and configures user data. On the source path, \texttt{sources\_set\_data} registers state via \texttt{MAM\_Data\_add}, while \texttt{targets\_retrieve\_data} reconstructs it with \texttt{MAM\_Data\_get\_pointer}, providing explicit hooks for data redistribution. The \emph{check} phase occurs at synchronization points where \texttt{MAM\_Checkpoint} (e.g., with \texttt{MAM\_WAIT\_COMPLETION}) tests for and waits on pending reconfigurations. The \emph{reconfiguration} phase uses MaM’s merge and migration mechanisms so that control resumes in \texttt{main\_loop} with an updated process set and redistributed state. A CI ecosystem for Proteo should integrate MaM’s routines with the underlying MPICH runtime and a resource manager to drive automated tests of source/target transitions, merges, and different resize patterns.

Across these solutions, a common structure emerges: each framework exposes a small set of primitives that implement a malleability workflow with three recurring concerns---initialization of a dynamic execution context, checks or handoff points where reconfiguration may be triggered, and reconfiguration mechanisms that alter the process layout and redistribute data. Despite differences in APIs, runtimes, and resource managers, these patterns suggest that a shared testing strategy is possible: one that exercises the same logical phases while instantiating framework-specific calls and deployment choices.

This observation motivates the central goal of this work: to distill these recurring patterns into a systematic space of test types and to support them with an HPC-oriented CI ecosystem that can be adapted across DynRM frameworks. The following sections introduce this test structure and CI environment, and later demonstrate how they can be instantiated concretely in a representative malleability framework.

\section{Related Work}\label{sec:related}


Several works explore continuous integration and testing practices for scientific software and HPC environments. 
Early efforts propose CI workflows that use container images and facility-managed runners to execute tests on HPC systems, for example combining Jenkins with Singularity containers and Slurm allocations to validate application behavior on DOE supercomputers~\cite{sampedro_continuous_2018}. 
More recent work introduces generic CI workflows tailored to HPC that orchestrate containers, HPC job launchers, and numerical result comparison tools within automated pipelines~\cite{society_of_research_software_engineering_building_2025,peters_enhancing_2025}.
In the broader research software engineering community, RSE-oriented workflows emphasize testing, CI/CD, and artifact linking as key enablers of sustainable, reproducible scientific software~\cite{maric_research_2022,schubert_promoting_2024}. 
These contributions show the feasibility and benefits of CI for scientific and HPC software, but they largely target applications and workflows rather than the specific multi-layer behavior of DynRM tools.

DynRM solutions have been studied extensively as a way to overcome the limitations of static batch scheduling in HPC systems. Traditional schedulers allocate a fixed amount of resources to each job for its entire lifetime, which leads to inefficiencies under fluctuating workloads and limits responsiveness to changing system conditions. In contrast, DynRM frameworks allow running jobs to expand or shrink their resource footprint at runtime in collaboration with the cluster Resource Management System (RMS), adapting to system load, energy constraints, or application progress~\cite{aliaga_survey_2022,tarraf_malleability_2024}. 
However, these studies primarily focus on mechanisms, performance, and scheduling effectiveness, but say little about systematic testing or continuous integration for such frameworks.

In this regards, several authors have proposed virtualized environments and container-based testbeds which help to reduce the cost of experimentation with complex runtime and resource-management stacks.
Hursey \textit{et al.} introduce a Docker-based virtual cluster targeting Open MPI, PRRTE, and OpenPMIx, providing a controlled and reproducible environment for functional and integration testing without direct access to production-scale systems~\cite{hursey2025pmix_swarm_toy_box}. 
A similar containerized approach has been adopted to support DynRM with the DPP abstractions, enabling developers to explore dynamic behaviors in a self-contained setting~\cite{dynres2025docker_cluster}. 
These environments significantly lower the barrier to experimentation and continuous development, yet their workflows are typically triggered manually and do not provide systematic, version-aware CI pipelines or test suites specialized for DynRM semantics.

In summary, there remains a gap in applying automated testing and CI specifically to DynRM frameworks, where correctness hinges on complex interactions among state machines, schedulers, and process managers. This work addresses this gap by presenting a tests taxonomy and a CI ecosystem, and their application in DMR aims to provide reproducible, version-aware validation of dynamic behavior and to support sustainable development of DynRM software.

\section{Methodology}\label{sec:methodology}
This section introduces the general approach followed in this work and provides the foundation for subsequent evaluation and discussion. First, we present a taxonomy of tests for MPI malleable libraries, which organizes the space of malleability-related tests along well-defined dimensions and clarify the distinction between functional and non-functional assessment. Then, we describe an HPC-oriented continuous integration ecosystem for DynRM, showing how this taxonomy can be instantiated in practice to automate malleability testing across different frameworks and runtime environments.

\subsection{Taxonomy of Tests for MPI Malleable Libraries}\label{subsec:taxonomy}
The test suite for MPI malleable libraries is structured around two main categories: \emph{component integration testing} and \emph{system testing}. Additionally, tests in the \emph{system testing} category are further classified as \emph{functional} or \emph{non-functional}, according to the type of behavior they assess.

\subsubsection{Component Integration Testing}\label{subsubsec:component_test}
Component integration tests are organized by purpose, following the standard malleability workflow: initialization, check, and reconfiguration. Unlike traditional unit tests, they exercise real dependencies---parallel runtimes and resource managers---rather than mocks. Instead of relying on isolated unit testing, we deliberately adopt integration-oriented tests, because the tight coupling typical of malleable systems it results impractical to test only the application logic in isolation from its underlying runtime and scheduling components.

In this regard, the test suite enumerates the key behaviors that must be validated at each phase of the workflow, as follows:

\paragraph{Initialization.}
Tests validate both state and environmental preconditions before enabling dynamic behavior:
\begin{itemize}
    \item \emph{Environment and dependency validation}: Initialization confirms all required runtime dependencies are available and verifies that library dependencies meet minimum version requirements.
    \item \emph{State validation}: The library rejects initialization on an invalid state, verifies that all required preconditions are satisfied, and guarantees that the system remains in a coherent state if initialization completes successfully.
    \item \emph{Configuration and argument handling}: Valid arguments are parsed correctly, while invalid arguments are handled gracefully with appropriate error reporting.
\end{itemize}

\paragraph{Check.}
The check phase validates readiness for potential reconfigurations while enforcing safety constraints:
\begin{itemize}
    \item \emph{State validation}: Checks reject invalid library states. 
    \item \emph{Guard \& inhibition mechanisms}: Checks respect active inhibition rules to prevent premature reconfigurations.
    \item \emph{No-op behavior}: Checks do not trigger reconfiguration when none is requested.
    \item \emph{Constraint validation}: Checks validate resource availability constraints, enforcing minimum and maximum resource limits.
    \item \emph{Request lifecycle management}: Triggered requests are processed correctly; pending requests are recognized; requests with timeouts handle expiration appropriately; and running requests report their state accurately.
\end{itemize}

\paragraph{Reconfigure.}
Tests ensure safe dynamic transitions:

\begin{itemize}
    \item \emph{State validation}: Reconfiguration validates preconditions and rejects invalid states.
    \item \emph{Integrity}: Jobs in the RMS queues remain unaffected unless explicitly marked to shrink or terminate.
    \item \emph{Scale transition handling}: Shrink operations execute correctly, and expansion operations complete successfully.
\end{itemize}

These tests are organized into three groups corresponding to reconfiguration stages:
\begin{itemize}
    \item \textbf{Resource reallocation}: New nodes are detected correctly.
    \item \textbf{Process layout reshape}: New processes are accessible via communicators.
    \item \textbf{Data redistribution}: Mechanisms enabling data redistribution are triggered in the correct order.
\end{itemize}

\subsubsection{System Testing}\label{subsubsec:system_test}
These tests exercise the full stack: application, malleability library, MPI, and resource manager.

\paragraph{Functional tests.}

Functional system tests focus on realistic scenarios where malleability is actually applied. Complete end--to--end tests verify that the system behaves correctly under representative evolution patterns:
\begin{itemize}
    \item The system evolves linearly from minimum to maximum scale and then back to the starting point, following a simple pattern. This exercise covers common \textbf{development} use cases in which applications grow and shrink gradually around a target size, and validates that repeated resize cycles do not accumulate inconsistent state.
    \item The system handles arbitrary, non-linear evolution scenarios. These scenarios more closely resemble \textbf{production} use cases, including alternation between growth and shrink phases and sequences that combine large and small resize steps.
\end{itemize}

\paragraph{Non-functional tests.}
The proposed non-functional system tests focus on assessing the quality of what we consider the core DynRM attributes: robustness, latency, and scalability. These attributes are broadly applicable to DynRM, in contrast to other common non-functional concerns such as performance (DynRM solutions are not necessarily designed to increase FLOPS), usability (they are not primarily about syntactic convenience), or reproducibility (which is often unattainable in real production environments).

This suite is limited by the fact that the overhead introduced by malleability depends heavily on the underlying network and the volume of data exchanged, and is therefore better characterized within each specific application study rather than through a generic test suite.  As a result, we limit the tests to areas involving dynamic resource manager libraries, aiming to capture indicative system behavior while avoiding exhaustive testing. These are as follows:

\begin{itemize}
    \item \textbf{Scalability:} The system is evaluated to determine its behavior up to a maximum arbitrary limit. By running these tests, we assess scalability by verifying the upper limit and robustness through repeated testing across multiple configurations within a short time window.
    \item \textbf{Time constraint:} The system is tested to ensure that a reconfiguration (without data transfer) completes within a predefined time limit. This measures that the latency of the resize operation remains below the specified threshold.
\end{itemize}

\subsection{Dynamic Resource Management CI Ecosystem}\label{subsec:ecosystem}
We deployed the CI workflow for HPC DynRM in a reproducible and portable environment that can run on generic servers or cloud infrastructure. This design enables the straightforward connection of the test suite to different CI services while preserving identical behavior across platforms. The current configuration relies on a CI pipeline executed by Jenkins\footnote{\url{https://www.jenkins.io}}, integrated with a GitLab\footnote{\url{https://www.gitlab.com}}-hosted repository so that tests are automatically triggered on every push and merge request, always validating the latest commit. Pipelines are described in Groovy\footnote{\url{https://groovy-lang.org}}, a Turing-complete language running on the Java Virtual Machine (JVM), which offers greater flexibility for expressing complex control flow, conditional execution, and dynamic test selection than purely declarative YAML-based configurations.

\begin{figure}
    \centering
    \includegraphics[width=0.95\linewidth]{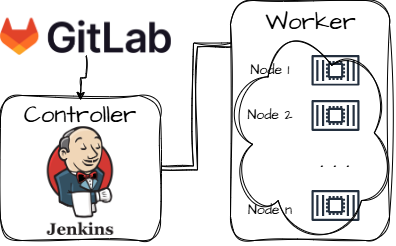}
    \caption{Jenkins deployment using Docker Compose. The \textit{controller} container runs the Jenkins server and is the only externally exposed component, while the \textit{worker} container, instantiated from a Docker-in-Docker (DinD) image, executes the containerized cluster-based CI pipelines.}
    \label{fig:jenkis-deploy}
\end{figure}

Figure~\ref{fig:jenkis-deploy} depicts the deployment, where a \textit{controller} runs the Jenkins server itself; it is the only component exposed externally and is responsible for triggering pipelines, managing credentials, and storing the global configuration of the CI environment.
Additionally, a \textit{worker} is responsible for executing the pipelines.
By offloading user jobs to this worker, we reduce computational load in the \textit{controller} and mitigate security risks.

\section{Applying the Testing Methodology: The DMR Use Case}
\label{sec:implementation}
This section demonstrates how the testing methodology and CI ecosystem---motivated by the recurring malleability patterns identified in Section~\ref{sec:motivation} and formalized in Section~\ref{sec:methodology}---can be instantiated in a concrete MPI malleability framework. We use the DMR framework as a representative case study that exercises the full range of component-integration and system-level tests across realistic HPC environments.

\subsection{The DMR Framework}
The Dynamic Management of Resources (DMR, pronounced “dimmer”) framework is a middleware layer that enables process malleability and dynamic resource management in MPI applications running on HPC systems~\cite{iserte_agut_high-throughput_2018}. Its goal is to let a running job acquire, release, or reconfigure compute resources at runtime---expanding or shrinking transparently during execution---to improve system utilization and responsiveness without user intervention.

DMR sits in between four main components: the scientific application, the MPI runtime, the performance monitor, and the RMS. Through this position, applications can query resource availability, request reconfigurations, and adapt their execution footprint dynamically, while DMR propagates these changes consistently to both the scheduler and the MPI process manager.

\begin{figure}
    \centering
    \includegraphics[clip,width=0.85\linewidth,trim={1cm 1cm 1cm 1cm}]{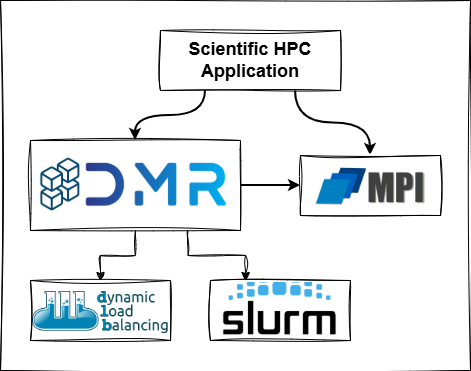}
    \caption{DMR architecture.}
    \label{fig:dmr}
\end{figure}

Figure~\ref{fig:dmr} illustrates how DMR’s components interact with the application. The MPI-enabled application invokes DMR routines, which themselves rely on MPI, and DMR in turn connects to the DLB performance monitor to retrieve runtime metrics and to the Slurm batch system to adjust resources on the fly.

Internally, DMR follows a state-driven architecture that mirrors the malleability workflow used in our taxonomy. A small set of API routines governs the main phases:
\begin{itemize}
  \item \texttt{dmr\_init}: Initializes the dynamic environment, configures communication with the RMS, and prepares internal data structures.
  \item \texttt{dmr\_check}: Invoked at synchronization points to indicate readiness for a potential resizing operation; DMR evaluates policies and system feedback to decide whether to reconfigure.
  \item \texttt{dmr\_reconfigure}: Executes the chosen reconfiguration and coordinates with the MPI runtime to spawn or terminate processes as needed.
  \item \texttt{dmr\_finalize}: Cleans up the management layer, ensuring that resources and communication contexts are properly released.
\end{itemize}

These routines implement a feedback-driven control loop in which the application, DMR, and the RMS interact continuously: when the application calls \texttt{dmr\_check}, DMR--- in coordination with the RMS---determines whether the current allocation should be expanded, shrunk, or left unchanged; if a resize is requested, it triggers the MPI runtime to integrate or remove processes, orchestrates any required data redistribution, and then resumes execution from the last synchronization point.

This multi-layer integration makes DMR a representative and demanding testbed for the proposed methodology. Small changes in DMR’s state machine, scheduler interaction, or process-management logic can propagate subtle effects across the stack, which motivates the use of a structured test taxonomy and an automated CI ecosystem to ensure correctness and maintainability as the framework and its ecosystem evolve.

\subsection{DMR within the Testing Taxonomy}\label{subsec:dmr_taxonomy}
\begin{figure*}
    \centering
    \includegraphics[width=1\linewidth]{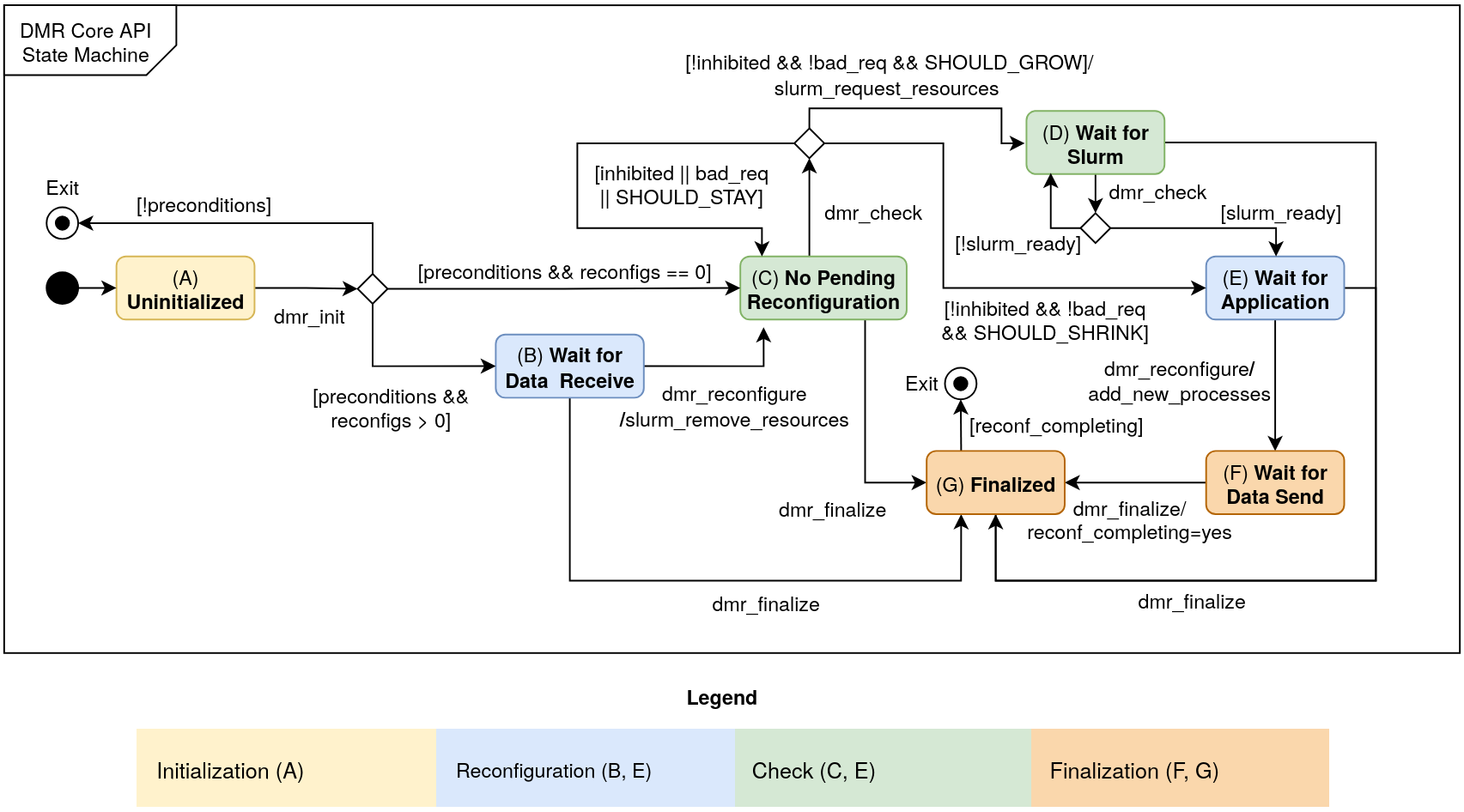}
    \caption{DMR Core API State Diagram.}
    \label{fig:dmr-state-diagram}
\end{figure*}

The DMR library can be modeled as a state machine in which execution progresses through a set of well-defined states in response to events, primarily invocations of its core API routines. 
Figure~\ref{fig:dmr-state-diagram} summarizes this behavior as a high-level contract between the application and DMR, abstracting implementation details while highlighting the most relevant states and transitions for testing. Most states include a transition to the \texttt{Finalized} state, allowing DMR to terminate at any point after initialization if required by the application.

In designing tests for DMR in the \textit{initialization} category of the component testing taxonomy (described in Section~\ref{subsubsec:component_test}), we focus on the behavior of the \texttt{Uninitialized} state shown in Figure~\ref{fig:dmr-state-diagram}. The only valid transition out of \texttt{Uninitialized} is a call to \texttt{dmr\_init}, and, conversely, this is the only state in which \texttt{dmr\_init} can be invoked. Therefore, we implement \textit{state validation} tests that check that \texttt{dmr\_init} rejects calls issued from any other state and that the expected transitions from \texttt{Uninitialized} to \texttt{Wait for Data Receive} and \texttt{No Pending Reconfiguration} are exercised correctly. 
Furthermore, DMR must validate a number of internal and external preconditions, represented in Figure~\ref{fig:dmr-state-diagram} as \texttt{preconditions}. If any of these preconditions is not satisfied at initialization time---for example, if an empty program name is provided---the expected outcome is that the library exits instead of proceeding in an inconsistent state. The full list of component integration tests implemented in DMR that relate to the \textit{Initialization} category and their position in the testing taxonomy is provided in Table~\ref{tab:initialization-tests}. 

\begin{table*}[]
\caption{DMR specific tests within the component integration testing \textit{Initialization} category. Similar tests are merged into a single description for brevity.}
\label{tab:initialization-tests}
\centering
\renewcommand{\arraystretch}{1.75}%
\begin{tabular}{| P{0.11\textwidth} | P{0.24\textwidth} | P{0.57\textwidth} |}
\hline
\rowcolor[HTML]{C0C0C0} 
\textbf{Category} & 
\textbf{Representation in Figure \ref{fig:dmr-state-diagram}} & 
\textbf{Description}                    
\\ \hline
State validation & 
The action \texttt{dmr\_init} is enabled exclusively in state A. & 
\texttt{dmr\_init} rejects calls to it when DMR is in any other state than \texttt{Uninitialized}.
\\ \hline
State validation &
Transition from state A to C. &
The library enters a standby state (\texttt{No Pending Reconfiguration}) when initialized for the first time, i.e. when the number of reconfigurations is 0. 
\\ \hline
State validation &
Transition from state A to B. &
The library enters a reconfiguration completion state (\texttt{Wait for Data Receive}) when \texttt{dmr\_init} is called and the number of reconfigurations is greater than 0.

\\ \hline

\multirow[c]{2}{*}{%
  \parbox{0.11\textwidth}{Environment and dependency validation}
} &
\multirow[c]{2}{*}{\parbox{0.22\textwidth}{The ``preconditions'' guard of \texttt{dmr\_init}.}} & 
Initialization detects if no Slurm job with a valid job ID is found in the environment.
\\ \cline{3-3}
& &
Missing DMR-specific environment variables are detected by \texttt{dmr\_init}.
\\ \hline 

\multirow[c]{3}{*}{\parbox{0.11\textwidth}{Configuration and argument handling}}&
\multirow[c]{3}{*}{\parbox{0.22\textwidth}{The ``preconditions'' guard of \texttt{dmr\_init}.}} & 
An argument count that is negative or zero is rejected; positive accepted.
\\ \cline{3-3}
&&
Argument array that is NULL or has NULL as its first element is rejected.
\\ \cline{3-3}
&&
An empty string passed as program name is rejected.   
\\ \hline

\end{tabular}
\end{table*}

States associated with the \textit{Check} category (described in Section~\ref{subsubsec:component_test}) in Figure~\ref{fig:dmr-state-diagram} are \texttt{No Pending Reconfiguration}---active while DMR is ready to accept new requests---and \texttt{Wait for Slurm}---where a submitted resource request has not yet been served by Slurm; from these states, transitions to all other states (except \texttt{Finalized}) are triggered by calls to \texttt{dmr\_check}. Given these constraints, we implement \textit{state validation} tests to ensure that \texttt{dmr\_check} returns an error when invoked outside the expected states. To transition from \texttt{No Pending Reconfiguration} to another state, we must also validate that the call is not inhibited (\texttt{inhibited} guard) and that the request is not invalid (\texttt{bad_req} guard, abbreviation of ``bad request''), thereby covering \textit{guard inhibition mechanisms} and \textit{constraint validation}, respectively. Finally, we verify that \textit{request lifecycle management} behaves correctly, ensuring that valid resource modification requests are served and that the \texttt{Wait for Slurm} state is exited once resources have been granted. The full list of component integration tests developed in DMR for the \textit{Check} category is provided in Table~\ref{tab:check-tests}.

\begin{table*}[]
\caption{DMR specific tests within the component integration testing \textit{Check} category. Similar tests are merged into a single description for brevity.}
\label{tab:check-tests}
\centering
\renewcommand{\arraystretch}{1.75}%
\begin{tabular}{| P{0.11\textwidth} | P{0.24\textwidth} | P{0.57\textwidth} |}
\hline
\rowcolor[HTML]{C0C0C0} 
\textbf{Category} & 
\textbf{Representation in Figure \ref{fig:dmr-state-diagram}} & 
\textbf{Description}                    
\\ \hline
State validation & 
The action \texttt{dmr\_check} is enabled exclusively in states C and D. & 
\texttt{dmr\_check} rejects calls to it when DMR is in any other state than \texttt{No Pending Reconfiguration} or \texttt{Wait for Slurm}.
\\ \hline
Guard \& inhibition mechanisms &
The "inhibited" guard of the \texttt{dmr\_check} action from state C. &
A call to \texttt{dmr\_check} does not trigger a reconfiguration if the current iteration is inhibited. If not inhibited, a reconfiguration is triggered.
\\ \hline
No-op behavior &
The "SHOULD_STAY" guard of the \texttt{dmr\_check} action from state C. &
Given a DMRSuggestion of \texttt{SHOULD\_STAY} (a no-op), the library does not begin to reconfigure.
\\ \hline
Constraint validation &
The "bad\_req" (bad request) guard of the \texttt{dmr\_check} action from state C. &
Attempting to shrink by a number exceeding the node count has no effect.
\\ \hline
Request lifecycle management &
Transition from state C to D. &
A valid request to add new resources is granted.
\\ \hline
Request lifecycle management &
Transition from state C to E. &
A valid request to remove resources is granted. 
\\ \hline
Request lifecycle management &
The "slurm_ready" guard of the \texttt{dmr\_check} action from state D. &
If a request for resources from Slurm is pending, the current state is maintained. If granted, the state progresses to \texttt{Wait for Application}.
\\ \hline

\end{tabular}
\end{table*}

Figure~\ref{fig:dmr-state-diagram} also shows the two states in which reconfiguration is driven through \texttt{dmr\_reconfigure}, namely \texttt{Wait for Data Send} and \texttt{Wait for Application}. Notice that \texttt{Wait for Data Send} is not part of DMR's \textit{Reconfiguration} component in that category (see Section~\ref{subsubsec:component_test}), since it belongs to application-specific data redistribution logic. As with the other components, we perform \textit{state validation} to ensure that \texttt{dmr\_reconfigure} has no effect when invoked outside \textit{Reconfiguration} stages, such as when DMR is \texttt{Uninitialized}. The DMR-specific component integration tests within the taxonomy at these stages are summarized in Table~\ref{tab:reconfigure-tests}. These tests primarily target correct addition and removal of resources, ensuring that \textit{integrity} is preserved---Slurm jobs not affected by a change in resources must remain unmodified---and that \textit{scale transition handling} completes correctly, for example by confirming that new processes are added and remain reachable over MPI.

\begin{table*}[]
\caption{DMR specific tests within the component integration testing \textit{Reconfigure} category. Similar tests are merged into a single description for brevity.}
\label{tab:reconfigure-tests}
\centering
\renewcommand{\arraystretch}{1.75}%
\begin{tabular}{| P{0.16\textwidth} | P{0.29\textwidth} | P{0.46\textwidth} |}
\hline
\rowcolor[HTML]{C0C0C0} 
\textbf{Category} & 
\textbf{Representation in Figure \ref{fig:dmr-state-diagram}} & 
\textbf{Description}                    
\\ \hline
State validation & 
The action \texttt{dmr\_reconfigure} is enabled exclusively in states B and E. & 
\texttt{dmr\_reconfigure} rejects calls to it when DMR is in any other state than \texttt{Wait for Data Receive} or \texttt{Wait for Application}.
\\ \hline
Integrity &
Action ``slurm\_remove\_resources'' in transition from B to C. &
Resources not marked for removal are not removed in reconfiguration.
\\ \hline
Scale transition handling (resource reallocation) &
Action ``slurm\_remove\_resources'' in transition from state B to C. &
Slurm jobs that are marked for removal are removed; Slurm jobs marked to shrink have resources removed.
\\ \hline
Scale transition handling (process layout re-shape) &
Action ``add\_new\_processes'' in transition from state E to F. &
The execution adds a new process when requested to and communication with it is established.
\\ \hline
Scale transition handling (process layout re-shape) &
Action ``add\_new\_processes'' in transition from state E to F. &
The execution adds a new process when requested to and writes an internal DMR checkpoint file to disk.
\\ \hline
\end{tabular}
\end{table*}

For system testing (see Section~\ref{subsubsec:system_test}), we implement a series of functional and non-functional tests for DMR.

The developed component integration tests provide a fast way to assess the overall health of the library and offer reasonable confidence in its correct behavior. However, because they run in relative isolation and do not exercise full system coordination, they may still miss issues that only appear after multiple reconfigurations or in more complex scenarios. To address this, system tests are used to provide stronger guarantees that the library behaves correctly under common conditions and edge cases. The functional system tests developed for DMR are organized into three categories:
\begin{itemize}
\item Manual resize tests
\item Policy resize tests
\item Data redistribution tests
\end{itemize}

\paragraph{Manual resize tests:} Designed to verify that the system as a whole can, over multiple iterations, reconfigure correctly at the granularity of MPI processes, compute nodes, and Slurm jobs, including scenarios that combine job termination and job shrinking. For example, assuming one process per node, the manual resize test \texttt{test_nonlinear_proc_reconfig} follows the pattern:
\begin{lstlisting}[numbers=none]
R0: J0[p0]
R1: J0[p0], J1[p1-p2]
R2: J0[p0], J1[p1-p2], J2[p3-p6]
R3: J0[p0], J1[p1-p2], J2[p3-p6], J3[p7-p9]
R4: J0[p0], J1[p1-p2], J2[p3]
R5: J0[p0]
\end{lstlisting}

The test starts with a single MPI process (initial configuration (\texttt{R0})) and expands to three processes in the first reconfiguration (\texttt{R1}) by adding two processes in a new job (\texttt{J1}). At reconfigurations \texttt{R2} and \texttt{R3}, it expands further by four and three processes, respectively, with jobs \texttt{J2} and \texttt{J3}. At \texttt{R4}, the execution is shrunk by six processes across two jobs (\texttt{J2} and \texttt{J3}), terminating \texttt{J3} and reducing \texttt{J2} to a single process. 
Finally, at \texttt{R5}, a shrink operation removes all processes in \texttt{J1} and \texttt{J2}. This test exercises the library’s ability to add varying numbers of processes across reconfigurations and to mark Slurm jobs for both termination and partial reduction within the same reconfiguration step.

\paragraph{Policy resize tests:}
These tests verify that decisions to resize the execution can be performed automatically. For example, DMR can target a specific communication efficiency,\footnote{\url{https://pop-coe.eu/node/69}} adjusting the allocated resources to match a desired efficiency level. Building on the validation of the manual resize tests, this category checks that DMR policies reconfigure the execution in the expected direction under predetermined conditions.

\paragraph{Data redistribution tests:}
These tests ensure that the data redistribution mechanisms are exposed and usable under various reconfiguration patterns. For example, the \texttt{test\_cr\_file\_writeread} test behaves as a simple DMR-based application that performs a single reconfiguration, writing a checkpoint file to disk and reading it back after restart.

In addition to the functional tests, DMR also includes non-functional tests that evaluate desired properties of a dynamic resource manager beyond basic system integration. For example, the \texttt{test\_reconf\_time} test verifies that no reconfiguration lasts longer than five seconds, where the reconfiguration time is measured from the moment the additional resources are secured until the application resumes after the process resizing completes.

\subsection{CI Pipeline Configuration for DMR}\label{subsec:dmr_pipeline}
The interactions between DMR and its main dependencies, namely the resource manager and the parallel distributed runtime (Slurm and Open MPI in this work), are complex and often constrained in large-scale shared systems. To properly exercise DMR’s orchestration role, these components must be present in the test environment, even if this increases configuration complexity. To obtain a quick, reproducible, and fully automated configuration while avoiding limitations such as resource contention, we deploy a Docker Compose-based virtual cluster that emulates the dependencies required by DMR, minimizing, in turn, the real resources consumed by the tests. The cluster is primarily designed to run Slurm and allows an arbitrary Slurm version to be selected by specifying a container tag. Building on an existing Dockerized Slurm cluster implementation~\cite{slurm_docker_cluster}, we extend it with features tailored to DynRM tests, including support for arbitrary node counts at launch time and passwordless ssh among nodes. In addition, we install the Check\footnote{\url{https://libcheck.github.io/check/}} framework inside the containerized environment to execute the \textit{component integration tests}.

\section{Experimental Results}\label{sec:evaluation}
This section evaluates the proposed testing methodology and CI ecosystem when instantiated for the DMR framework. First, we validate that DMR and its test suite operate correctly on a production-scale supercomputer, demonstrating that the approach is compatible with realistic HPC environments. Then, we assess the behavior of the same configuration across multiple Slurm releases, highlighting how the CI pipeline exposes version-specific incompatibilities and supports systematic dependency management.

\subsection{Validation in a Supercomputer}
MareNostrum~5 (MN5) is a pre-exascale system integrated into the EuroHPC-JU infrastructure.\footnote{\url{https://bsc.es/marenostrum/marenostrum-5}}
Its general-purpose partition (GPP)\footnote{\url{https://www.bsc.es/supportkc/docs/MareNostrum5/overview/\#marenostrum-5-gpp-general-purpose-partition}} 
comprises 6{,}192 Intel Sapphire Rapids-based nodes, each equipped with two 
Intel Xeon Platinum 8480+ processors (56~cores at 2~GHz, 112~cores total) and 256~GB of DDR5 memory, interconnected by a 100~Gbit/s ConnectX-7 NDR200 InfiniBand network and managed by Slurm~23.02.7.

To validate the proposed approach, the containerized environment is configured to mirror MN5 by using the same Slurm version. In the same CI pipeline, DMR is also validated against the custom resource manager Slurm4DMR,\footnote{\url{https://gitlab.bsc.es/accelcom/releases/dmr/tools/slurm4dmr}} which is typically deployed by DMR application developers on a subset of MN5 nodes to provide additional flexibility beyond the standard RMS. The pipeline runs tests against both Slurm versions because their API differences are significant and compiling against them exercises distinct code paths in DMR.  Using the virtual cluster, each test specifies the Slurm version to use and the node count required for its execution.
We deliberately avoid validating all supported Slurm releases concurrently in order to reduce computational load and to mirror realistic deployment settings. In practice, and based on our experience, CI pipelines are best configured per platform, with each pipeline targeting the specific Slurm version of the platform (i.e., Slurm~23.02.7 for DMR in MN5). While DynRM \textit{library} developers may need to track compatibility across several Slurm versions, it is neither realistic nor necessary to support the full historical and bleeding-edge spectrum; site operators and application developers usually focus on a single, well-defined Slurm release in their production environment.

\begin{figure}
    \centering
    \includegraphics[width=0.75\linewidth]{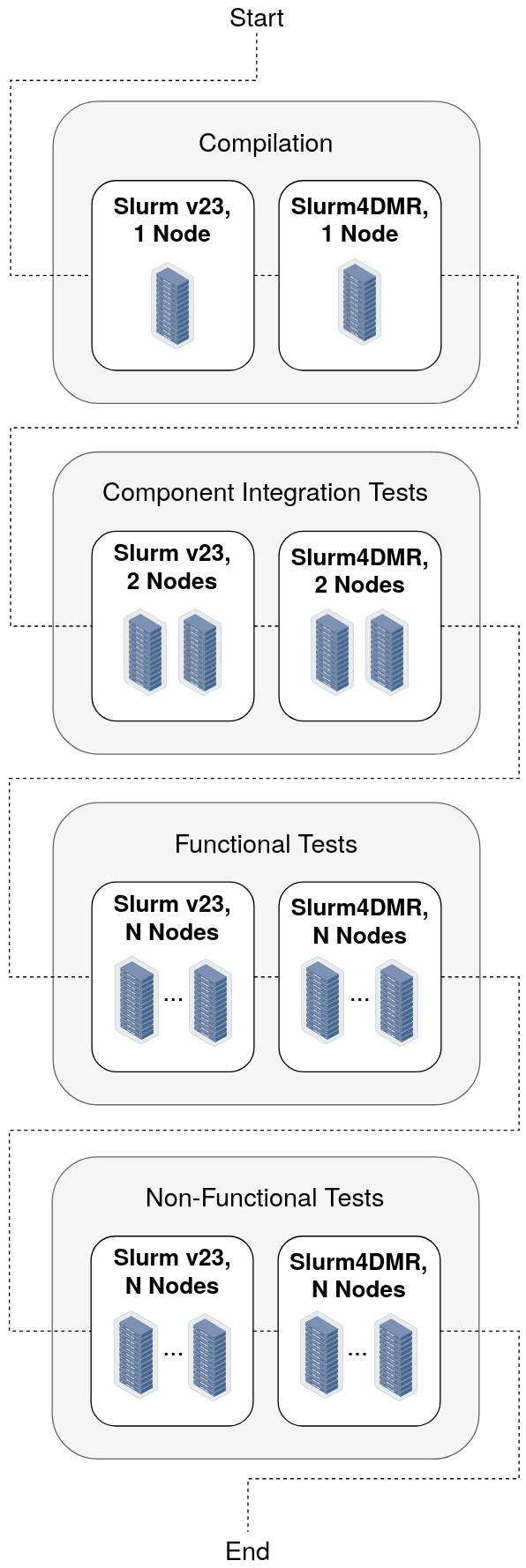}
    \caption{DMR's CI pipeline set up with Slurm version 23.02.07 (shown as Slurm v23) and the DMR-specific resource manager Slurm4DMR. The dotted line shows the path through the pipeline when all stages complete successfully.}
    \label{fig:dmr-pipeline}
\end{figure}

Figure~\ref{fig:dmr-pipeline} illustrates the CI pipeline, which is triggered by external events such as new merge requests.
First, DMR is compiled against both the target Slurm version and DMR's malleability-enabled Slurm version, inside their respective containerized environments; since compilation only requires a single process, the virtual clusters are launched with one node. 
After successful compilation, the \textit{component integration tests} are executed; these do not involve full system coordination and complete relatively quickly for any Slurm version, so they provide early detection of major issues. 
A two-node virtual cluster is deployed to run these tests, which require exactly two nodes.

Once the component integration tests finish, the more time-consuming \textit{functional tests} are run. 
Because these tests have different node-count requirements, multiple virtual cluster instances are launched with the appropriate size for each test. 
An alternative would be to run all tests in a single virtual cluster configured with the maximum required node count, but the per-test clusters provide better isolation, since components such as Slurm and MPI can leave unwanted side effects if not reset in between tests. 
After the general stability of the code has been verified, the \textit{non-functional tests} confirm that DMR, compiled against each Slurm version under evaluation, meets the relevant non-functional requirements at the node scale required by each test.

After a complete run of the pipeline in our containerized environment, we expect most issues that could arise relating issues or incompatibilities with the target platform's Slurm version, 23.02.07 on MN5, and Slurm4DMR, to have been detected. To verify this, we run the tests outside of the pipeline directly on the target platform outside of the containerized environment, using both evaluated Slurm versions. Running the same tests in the MN5 production environment lasted, in this case, 1 hour 32 minutes and 30 seconds, passing all stages without errors. The vast majority of this time is a result of queuing for additional resources, since in this environment there is contention for resources. Note that this waiting time depends on the current load of the shared cluster, and it is therefore uncontrolled and stochastic. In contrast, the Slurm4DMR tests completed in 3 minutes and 30 seconds, since these tests rely on pre-reserved resources and therefore any waiting time due to resource contention is already complete at the time the tests start.

\subsection{Evaluation Across Multiple Slurm Versions}
As introduced earlier, the containerized environment is configured to support multiple Slurm releases, enabling a systematic assessment of version compatibility for DMR and other malleability frameworks. In this study, the latest major Slurm versions from 17 onwards are evaluated, with the corresponding CI outcomes summarized in Table~\ref{tab:slurm_versions}. Additionally, the malleability-enabled Slurm variant Slurm4DMR is included in the evaluation (first row).

For each version, the table reports a short descriptive note (in parentheses after the version), the CI outcome (column ``Status''), and either the reason for failure or the total execution time. While runtime is not critical for these experiments, differences in execution time across versions reveal distinct behaviors and internal changes in Slurm's resource management. 
It is also important to note that the customized Slurm4DMR version distributed with DMR skips job-management tests, since it resizes jobs directly rather than launching and joining additional jobs, and its times are therefore not directly comparable. 

All versions from 20 to 23 completed successfully, consistent with the fact that DMR is already deployed and exercised in production environments using Slurm~23.02.7-1 (MN5), 23.11.4-1 (QMIO\footnote{\url{https://cesga-docs.gitlab.io/qmio-user-guide}}), and 23.11.11-1 (Leonardo\footnote{\url{https://www.hpc.cineca.it/systems/hardware/leonardo}}).  In contrast, for releases 17, 18, and 19, the pipeline execution failed because the tests assessing DMR functionality are based on scheduler APIs that are not compatible with these versions. In particular, the \textit{test\_check\_timeout\_expander}, \textit{test\_req\_grow}, \textit{test\_check\_pending\_job}, \textit{test\_kill\_running\_expander}, \textit{test\_shrink\_running\_expander}, and \textit{test\_no\_modify\_running\_expander} tests invoke the \texttt{slurm\_submit\_batch\_job} function, whose parameters differ from those in later Slurm releases. As DMR directly uses this Slurm functionality, this indicates it would also fail at the point where this function is called. For Slurm~24.11.5-1 and 25.11.2-1, the pipeline failed at compile time, indicating API or build-system changes that require code adaptation.
These results show that an automated, version‑parametrized CI configuration can expose both behavioral and compilation-level incompatibilities early, providing developers with precise feedback on when and where framework updates are required.  Unlike other software (e.g., web applications), DynRM middleware is not expected to support every historical or bleeding‑edge dependency: Slurm upgrades in HPC facilities are relatively infrequent, and changes across versions may introduce non-trivial API and configuration differences.  Consequently, both developers and site operators typically target a single, or a small set of well-defined, Slurm versions deployed on their production systems rather than aiming for broad, version-agnostic support.

\begin{table*}[h]
  \centering
  \begin{tabular}{llr}
    \toprule
    \textbf{Version (Notes)}        & \textbf{Status}      & \textbf{Time/Reason} \\
    \midrule
    \textit{Slurm4DMR} (based on slurm-17.02.0-0pre1) & Passed & 264 seconds. \\
    \midrule
    slurm-17-11-13-2        & Failed        & 6 tests not passed   \\
    slurm-18-08-9-1         & Failed        & 6 tests not passed         \\
    slurm-19-05-8-1         & Failed   & 6 tests not passed\\
    slurm-20-11-9-1         & Passed                            & 341 seconds    \\
    slurm-21-08-8-2         & Passed                            & 305 seconds     \\
    slurm-22-05-11-1        & Passed                            & 321 seconds    \\
    slurm-23-02-7-1 (MN5)        & Passed                            & 304 seconds     \\
    slurm-23-11-4-1 (QMIO)        & Passed                            & 1200 seconds     \\
    slurm-23-11-11-1 (Leonardo)       & Passed                     & 1198 seconds \\
    slurm-24-11-5-1         & Failed & Compilation error \\
    slurm-25-11-2-1         & Failed     & Compilation error  \\
    \bottomrule
  \end{tabular}
  \caption{CI results for DMR across multiple Slurm versions. Showed times are not expected to be a performance metric but a validation metric.}
  \label{tab:slurm_versions}
\end{table*}

\section{Discussion}\label{sec:discussion}
The DMR case study provides initial evidence that a structured test space and an HPC-oriented CI ecosystem can be instantiated in a realistic DynRM framework and used to exercise both component-level and system-level malleability behavior. The successful application of the approach to DMR, including automated validation across multiple Slurm versions and on a production supercomputer, suggests that other DynRM solutions exposing analogous initialization, readiness, and reconfiguration primitives could adopt similar patterns with limited adaptation effort. At the same time, the specificity of DMR’s architecture and integration points highlights the need to carefully account for framework- and site-specific details when transferring the approach elsewhere.

Although the presented tests taxonomy and the CI ecosystem is framework-agnostic, the implementation of the CI ecosystem is currently tailored to containerized, Slurm-based environments and MPI runtimes that expose specific malleability or sessions features. This choice makes the system practical and reproducible, but it restricts direct applicability to platforms that use other resource managers, non-MPI programming models, or site-specific CI services. Second, the evaluation focuses on small to medium-scale virtual clusters and a single production system; while this configuration is representative of many development workflows, it does not fully capture the behavior of extreme-scale deployments or facility-wide CI integrations. Third, the test space is deliberately restricted to robustness, latency, and scalability of the DynRM mechanisms themselves, leaving aside application-specific performance, energy, and numerical validation, which must be handled by complementary test suites.

These limitations open several directions for future work. One avenue is to generalize the CI ecosystem to additional resource managers (i.e., PBS, OAR, Flux) and to non-MPI runtimes (i.e.,  asynchronous many-task (AMT), task-based), exploring how the proposed test structure extends to other forms of malleability.

\section{Conclusions}\label{sec:conclusions}
The work presented in this paper shows that dynamic resource management for HPC can be supported by a systematic, test-driven methodology rather than by ad hoc, manually curated experiments. By introducing a taxonomy of tests for MPI malleable libraries and coupling it with an HPC-oriented CI ecosystem, the approach enables structured validation of both functional and non-functional behavior at component and system level, while exercising real MPI runtimes and resource managers instead of isolated mocks. 

Instantiating the methodology in the DMR framework demonstrates its practicality in a realistic setting. The DMR test suite covers initialization, check, and reconfiguration phases through state-machine-driven component integration tests, as well as end--to--end system tests, and the CI pipelines validate these behaviors both on a production supercomputer (MareNostrum 5) and across a range of Slurm releases, where the version-parameterized setup exposes API incompatibilities and build issues early.  These results highlight concrete benefits for DynRM software, including earlier fault detection, safer dependency upgrades, and easier maintenance as runtimes and schedulers evolve. 


\section*{Acknowledgements}
Language polishing was performed using an AI language model, with subsequent thorough review and editing by the authors. The authors are fully responsible for the final manuscript content.

\section*{CRediT Authorship Contribution Statement}
\begin{itemize}
\item Petter Sandås: Software, Investigation, Writing - Original Draft
\item Íñigo Aréjula-Aísa: Software, Validation, Writing - Original Draft
\item Sergio Iserte: Conceptualization, Methodology, Writing - Original Draft, Writing - Review \& Editing, Supervision.
\item Antonio J. Peña: Writing - Review \& Editing, Resources, Funding acquisition, Supervision.
\end{itemize}


\section*{Funding Sources}
This project is co-funded by the Ministerio para la Transformación Digital y de la función pública, within the framework of the Plan de Recuperación Transformación y Resiliencia, and by the European Union – NextGenerationEU. The views and opinions expressed are solely those of the author(s) and do not necessarily reflect those of the European Union. Neither the European Union nor the European Commission can be held responsible for them.
Antonio J. Peña was partially supported by the Ramón y Cajal fellowship RYC2020-030054-I funded by MCIN\slash AEI\slash 10.13039\slash 501100011033 and by ``ESF Investing in your future''.

\section*{Data Availability}
DMR is open-source software and the release version can be downloaded from \url{https://gitlab.bsc.es/accelcom/releases/dmr} and \url{https://zenodo.org/records/17831722}.

\bibliographystyle{elsarticle-num}
\bibliography{bib/bib}

\end{document}